# A microwave plasma source for VUV atmospheric photochemistry

S Tigrine[1,2], N Carrasco[1,3], L Vettier[1] and G Cernogora[1]

[1] LATMOS, Université Versailles St. Quentin, UPMC Univ. Paris 06, CNRS, 11 Bvd. d'Alembert, 78280 Guyancourt, France,

[2] Synchrotron SOLEIL, l'Orme des Merisiers, St Aubin, BP 48, 91192 Gif sur Yvette Cedex, France,

[3] Institut Universitaire de France, 103 Boulevard St Michel, 75005 Paris, France

E-mail : sarah.tigrine@latmos.ipsl.fr

**Abstract.** Microwave plasma discharges working at low pressure are nowadays a well-developed technique mainly used to provide radiations at different wavelengths. The aim of this work is to show that those discharges are an efficient windowless VUV photon source for planetary atmospheric photochemistry experiments. To do this, we use a surfatron-type discharge with a neon gas flow in the mbar pressure range coupled to a photochemical reactor. Working in the VUV range allows to focus on nitrogen-dominated atmospheres (λ<100nm). The experimental setup makes sure that no other energy sources (electrons, metastable atoms) than the VUV photons interact with the reactive medium. Neon owns two resonance lines at 73.6 and 74.3 nm which behave differently regarding the pressure or power conditions. In parallel, the VUV photon flux emitted at 73.6 nm has been experimentally estimated in different conditions of pressure and power and varies in a large range between $2\times10^{13}$ ph.s$^{-1}$.cm$^{-2}$ and $4\times10^{14}$ ph.s$^{-1}$.cm$^{-2}$ which is comparable to a VUV synchrotron photon flux. Our first case study is the atmosphere of Titan and its $N_2$-$CH_4$ atmosphere. With this VUV source, the production of HCN and $C_2N_2$, two major Titan compounds, is detected, ensuring the suitability of the source for atmospheric photochemistry experiments.

**PACS.**

| | |
|---|---|
| 50 | Physics of gases, plasmas, and electric discharges |
| 52.70.Kz | Optical (ultraviolet, visible, infrared) measurements |
| 52.80.Pi | High-frequency and RF discharges |
| 52.80.Yr | Discharges for spectral sources (including inductively coupled plasma) |

1. Introduction

Microwave plasma columns are widely used this is why developing an effective source has become an important topic over the past decades. In the 1970s, Moisan et al (1)(2) created the first surface-wave launcher, called "surfatron", working in the micro wave range, which is an efficient and compact tool for creating long plasma columns in a wide range of pressures (from atmospheric pressure to fractions of mbars) and discharge diameters. One of the main advantages of those discharges is that they are electrodeless and then avoid pollution by metal sputtering. They are also reliable reservoirs of charged, reactive particles and radiation (3). The individual wavelength of these radiations depends on the gas used for the discharge. It goes from the infra-red up to the Vacuum Ultra-Violet (VUV) range (below 200 nm) as shown by (4) in their study of the VUV emission of an Ar-$H_2$ gas mixture.

The radiations are mostly used as a light source for different technical studies. For example, a low pressure Kr-VUV source is used for its photoionization efficiency, as an alternative to electronic impact in mass spectrometers, in order to avoid complex fragmentation patterns (5). At atmospheric pressure,

A microwave plasma source for VUV atmospheric photochemistry surface-wave driven plasma find applications in surface treatment, as they increase the wettability of silicon (6).

The aim of the present work is to develop a low-pressure (mbar) microwave plasma discharge (2.45 GHz) in a noble gas used as a VUV light source for a new application: to perform laboratory simulations of the planetary atmospheric photochemistry occurring at high altitudes, where solar VUV photons are penetrating. This wavelength range is of high demand in photochemistry laboratory simulations as it permits the dissociation and/or ionization of a lot of molecules, involving molecular nitrogen, the most abundant component in the atmospheres of the Earth and of Titan, the largest satellite of Saturn, on which we will focus here. As a matter of fact, thanks to the Cassini-Huygens mission (NASA-ESA), it is known now that Titan's atmosphere is mainly composed of $N_2$ and $CH_4$ for a methane concentration between 2 and 10%, depending on the altitude (7). The observations revealed that a complex organic chemistry takes place in this 1200km-wide atmosphere, all starting with the dissociation and ionization of the neutrals $N_2$ and $CH_4$ by the VUV solar radiations in the upper layers.

The fast and efficient photochemistry leads to the formation of heavy C and/or N-based species, like benzene ($C_6H_6$) or pyridine ($C_5H_5N$) for example. They will then grow into aerosols that globally surround the whole satellite on several hundreds of kilometres. All those processes make Titan a natural laboratory to witness and understand this complex organic chemistry but despite all the data collected, all the possible photochemical pathways in such a hydrocarbon-nitrogen-rich environment are not precisely understood.

This is why Titan's atmospheric photochemistry experiments are of high interest. So far, the main VUV sources have been Mercury lamps at 253.65 nm (8) or Hydrogen lamps emitting the $Ly_α$ line at 121.6 nm which is the main source of VUV energy in the solar system (9). Those lamps allow to work in closed gas cells as $MgF_2$ windows are transparent to radiations above 110 nm and focus on the kinetics of heavy hydrocarbon formation (10), or on the influence of benzene ($C_6H_6$) on Titan's atmospheric chemistry (11). Also, a gas mixture of $H_2$/He was used in various conditions in order to obtain an output comparable to the solar irradiance between 115 and 170 nm (12).

Nevertheless, molecular nitrogen requires wavelengths shorter than 100 nm in order to allow a direct photo-dissociation (Figure 1). This implies that the flows on both sides have to be balanced as the gas reactor and the VUV source would be openly coupled. So far, one way to obtain a VUV intense photon source has been coupling a gas reactor to a VUV synchrotron beamline ((13), (14)). But even if they allow a high photon flux and a large range of specific energies (15), they provide only a limited duration for the experimental campaigns as the demand has exponentially increased in the past decades.

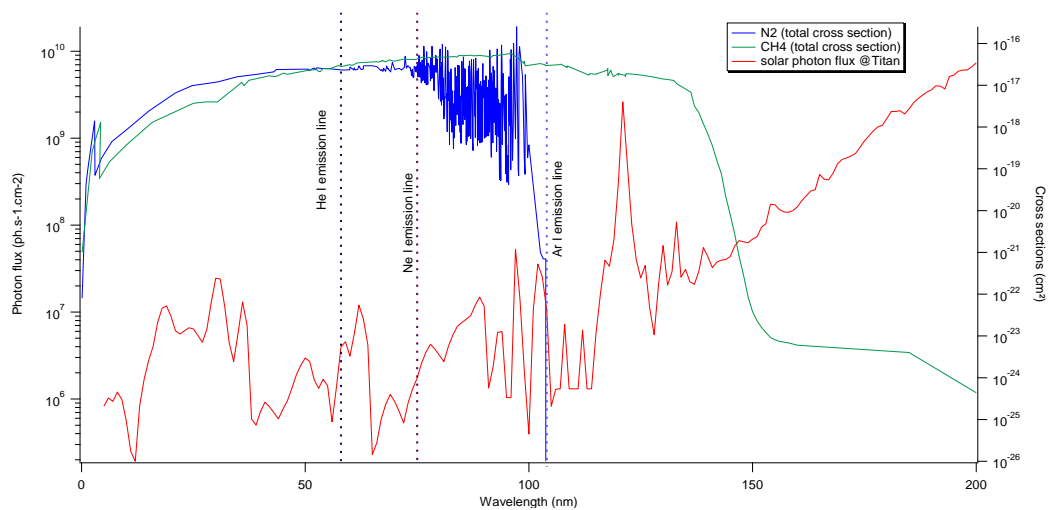

**Figure 1**: VUV absorption cross sections of $N_2$ (Blue) and $CH_4$ (Green), and the VUV solar spectrum at Titan (Red). The dotted vertical lines show the VUV emission lines of the three rare gases HeI, NeI, ArI.

# A microwave plasma source for VUV atmospheric photochemistry

Still in the VUV range, an alternative could be commercial lamps, which exist in different types and shapes. Most of them are based on plasma discharge techniques involving noble gases. A priori, they offer several experimental qualities like a high intensity VUV beam ($10^{15}$ ph.s$^{-1}$) or a low-pressure discharge (lower than few mbars). Nevertheless, these set-ups do not provide the flexibility needed in terms of wavelength and pressure that atmospheric photochemistry experiments demand. In fact, the Solar system offers a wide range of atmospheres which do not share the same conditions, especially regarding the pressures. The VUV source, which is to be window-less coupled to the photochemical reactor, has to suit an Earth-like stratospheric pressure as well as Titan's ionosphere conditions. For example, the Scienta VUV 5000 focuses only on the Extreme UV Helium lines at 58.4 nm, 53.7 nm (He I) and 30.4 nm (He II). Or the UVS40 A2 source from Henniker Scientific is designed to be only coupled with experiments at pressures lower than a few $10^{-2}$ mbars, which makes it inappropriate for some experimental simulations of atmospheric photochemistry.

For this kind of applications, new window-less VUV sources are required and flexible low-pressure microwave discharges could be a solution, explored in the present work. We test here the potential of the surfatron technique with our experimental setup. One of the main aims is to ensure that no other energy source than VUV radiations could reach the reactor and initiate some unwanted chemistry. The other energy sources to avoid are electrons from the plasma discharge and metastable atoms.

A complete diagnosis of the system is then performed regarding the flux of photons delivered by the VUV lamp in different conditions of pressure and power, in order to compare it with the flux of other VUV sources like synchrotron facilities. Conclusions regarding its suitability for pure atmospheric photochemistry experiments are then drawn, with the application to the upper atmosphere of Titan.

## 1. Experimental setup

### 1.1 The VUV source: a microwave-plasma discharge

The plasma discharge takes place in a 40-cm length quartz tube with an internal diameter of 8 mm and an external diameter of 10 mm, surrounded by a surfatron resonance cavity. The microwave power delivered by a Sairem generator goes up to 200 W. The surfatron can be moved along the quartz tube in order to settle the end of the discharge regarding the entrance of the photochemical reactor. A compressed air circulation avoids any over heat of the system. Surface temperature measurement showed that it does not go above 60°C.

The gas flow is regulated with a 0-10 sccm (cm$^3$.min$^{-1}$ STP) range MKS mass flow controller. Moreover, the pressure is measured with a Pfeiffer capacitor gauge mounted upstream the surfatron device while the main part of the plasma column is created downstream the field applicator.

Three noble gases, with intense resonance lines, can be used in order to work in the VUV range. Argon emits at 104.8 nm (no dissociation of $N_2$), Neon at 73.59 and 74.37 nm and He at 58.4 and 53.7 nm (ionization of both molecules) (table 1). Here, we use Neon: at this energy, the dissociation and ionization of $N_2$ are both possible (figure 1) in parallel to the ionization of $CH_4$ (threshold at 98.52 nm).

**Table 1.** Resonance lines of He I, Ne I and Ar I (NIST(16)).

| Gas | Wavelength (nm) | $A_{ki}$ (s$^{-1}$) |
| --- | --- | --- |
| He I | 58.43, 53.70 | $1.8 \times 10^9$ ; $5.66 \times 10^8$ |
| Ne I | 73.59 ; 74.37 | $6.11 \times 10^8$ ; $4.76 \times 10^7$ |
| Ar I | 104.8 ; 106.66 | $5.1 \times 10^8$ ; $1.19 \times 10^8$ |

### 1.2 Characterization of the source with a VUV monochromator

The characterization of the VUV source is performed by coupling it directly to a VUV 1-m focal-length McPherson NOVA 225 spectrometer, fitted with a 1200 gr/mm concave grating blazed at 45 nm and sensitive from 30 nm to 300 nm. The efficiency of the grating strongly varies with the wavelength and is around 6% at 75 nm (characterized by McPherson). In order to avoid the saturation of the acquisition system but still have a good resolution, the width of the slits is set up at 75 µm for a height of 4mm. The acquisition system consists of an Optodiode AXUV100 photodiode coupled to an amplifier. The output signal gives 1V for 1nA and saturates at 10V. Moreover, according to the data sheet, at 75 nm, the photodiode responsivity is about 0.22 A.W$^{-1}$.

The monochromator is pumped with a turbomolecular pump (Agilent TV 301) and the internal pressure is measured by a Penning gauge (Oerlikon) calibrated for air. The real pressure is obtained using a neon/air correction factor. The turbomolecular pump ensures an ultimate vacuum of 10$^{-8}$ mbar inside the monochromator; while the VUV source is also pumped by this system, through the entrance slit. At the maximum gas flow injected into the VUV source, the monochromator pressure is about 10$^{-3}$ mbar.

The luminous power collected by the photodiode is related to the number of photons per second:

$$\frac{dE}{dt} = \frac{hc}{\lambda} \times \frac{dN_p}{dt} \tag{1}$$

where E is the energy and N$_p$ the number of collected photons.

As neon from the source is pumped in the monochromator, optical absorption happens. Then the transmission inside the monochromator at a specific pressure is given by the Beer-Lambert law:

$$\frac{I_t}{I_0} = e^{-d \times [Ne] \times \sigma} \tag{2}$$

where d stands for the length of the optical path (here d=2 m); σ is the absorption cross section (for neon and around 75 nm, σ=9×10$^{-17}$ cm² (17)); and finally [Ne] is the neon density linked to the pressure via the perfect gas law. In the few cm between the end of the discharge and the monochromator entrance slit, the absorption is negligible.

After weighting the number of photons per second with all those factors (responses of both the photodiode and the grating, plus the neon absorption), we divide by the surface of the exposed slits in order to obtain the photon flux (ph.s$^{-1}$.cm$^{-2}$).

A microwave plasma source for VUV atmospheric photochemistry

*1.3  The atmospheric photoreactor ("APSIS")*

The table-top VUV source is to be coupled with a photochemical reactor named APSIS (Atmospheric Photochemistry SImulated by Synchrotron) in order to carry research on planetary upper atmospheres and their interaction with the VUV solar light. The APSIS reactor is described in details in (18). Briefly, it is a stainless steel chamber of dimensions 500 mm x 114 mm x 92 mm where the reactive mixture is introduced via a gas inlet (figure 2). The gas mixture is chosen in order to simulate Titan's atmospheric composition. In the present work, a 95-5% $N_2$-$CH_4$ mixture is injected up to 10 sccm with a MKS gas flow controller. Before each experiment, the reactor is pumped by a turbo-molecular pump down to $10^{-7}$ mbar. During the photochemistry experiments, a rotary vane pump ensures a stable pressure on the order of 1 mbar.

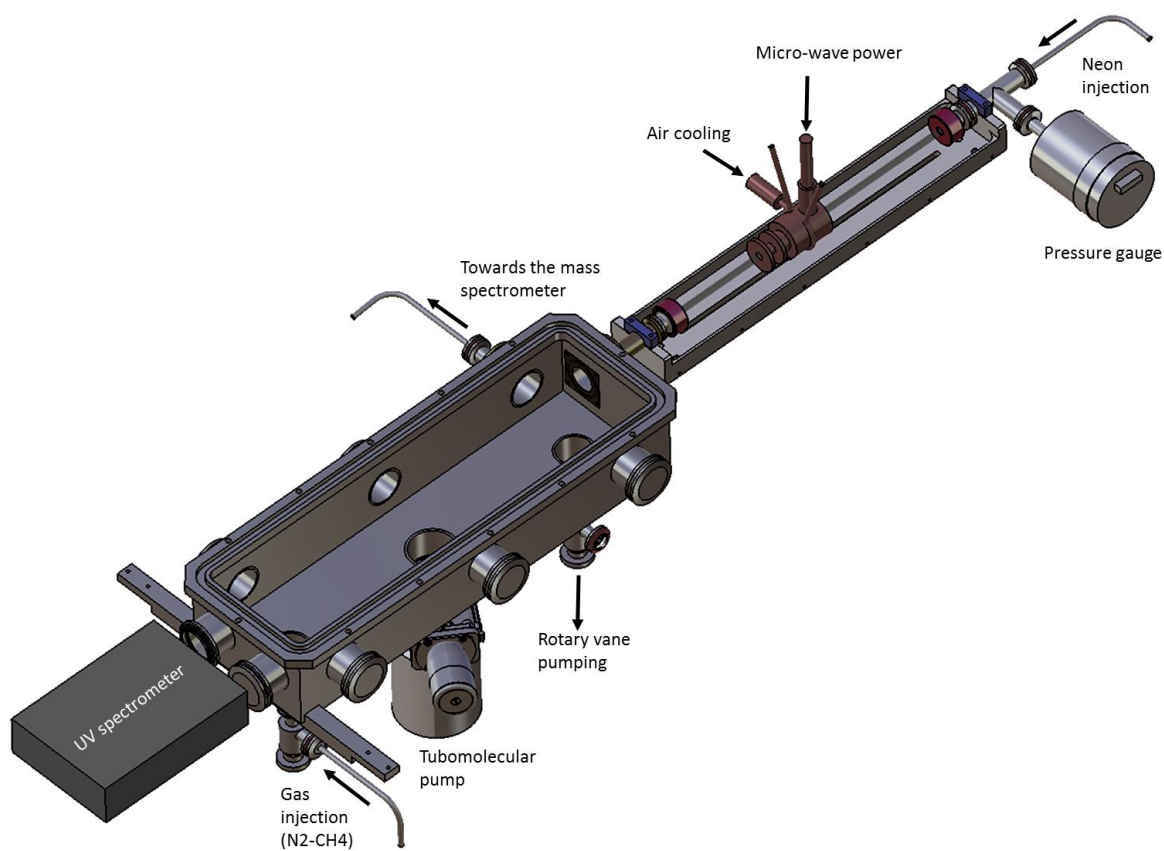

**Figure 2.** Scheme of the photochemical reactor coupled window-less with the VUV source.

*1.4  In-situ mass spectrometry on APSIS*

We use mass spectrometry to monitor the neutral molecules in the experiment, with a HIDEN HPR-20 QIC mass spectrometer. The gaseous products are taken at the closest spot from the VUV source. In order maintain a low pressure inside the mass spectrometer at $10^{-7}$ mbar, a 1m-capillary tube with an internal diameter of $1/16^e$ inch is used. The Multiple Ion Detection (MID) mode on the mass spectrometer is chosen, as it follows selected mass signal in function of time and shows the evolution of the intensities before and during the irradiation process.

*1.5  UV optical emission spectroscopy on the coupled system "VUV source-APSIS"*

In order to ensure that no neon plasma enters the APSIS reactor, and that the dissociation of nitrogen and methane only occurs from the VUV photochemistry, optical emission spectroscopy of the neon-$N_2$-$CH_4$ system is performed in the UV range. A Hamamatsu spectrometer TG-UV C9404CAH is positioned in front of the VUV source at the opposite side of the reactor (figure 2). The emitted spectra in the UV range from 200 nm to 450 nm is recorded through a quartz window.

*2.* **Results**

*2.1   Validation of pure photochemistry*

In order to work with a clean neon plasma, the discharge remains on for at least 20 min before starting any measurement. A VUV spectrum of the clean discharge is presented on figure 3 (black curve). The key parameter in our photochemistry experimental platform is the distance between the end of the discharge and the entrance of the reactor. Indeed, if the distance is too short, there is a risk that the discharge would continue inside the reactor and create a $N_2$-$CH_4$ plasma (electron-driven chemistry) or that the metastable neon atoms will enter the reactor. But if the distance is too long, a photon loss is to be considered.

*1.1.1   Preventing electron-driven chemistry.* Thanks to optical emission spectroscopy, it is possible to monitor the $N_2$ bands of the Second Positive System (SPS), which are recognizable because degraded to shorter wavelengths. The SPS is emitted from the $N_2(C^3\Pi_u)$ state which cannot be populated neither by the 73.6 nm nor the 74.3nm-VUV line. This emission is then due to the energetic electrons of the plasma in the reactor: seeing those SPS bands means that the plasma discharge enters the photo-reactor and interacts with the $N_2$-$CH_4$ mixture.

First, in order to inject the maximum amount of VUV photons, the surfatron has been placed as near as possible from the entrance of the APSIS reactor. However, in this configuration, $N_2$ bands of the SPS appear on the spectrum (Figure 3, red curve).

It is then relevant to balance the distance from the end of the discharge and the entrance of the reactor in order to have enough photons entering the reactive medium without exciting the nitrogen-methane mixture.

A microwave plasma source for VUV atmospheric photochemistry

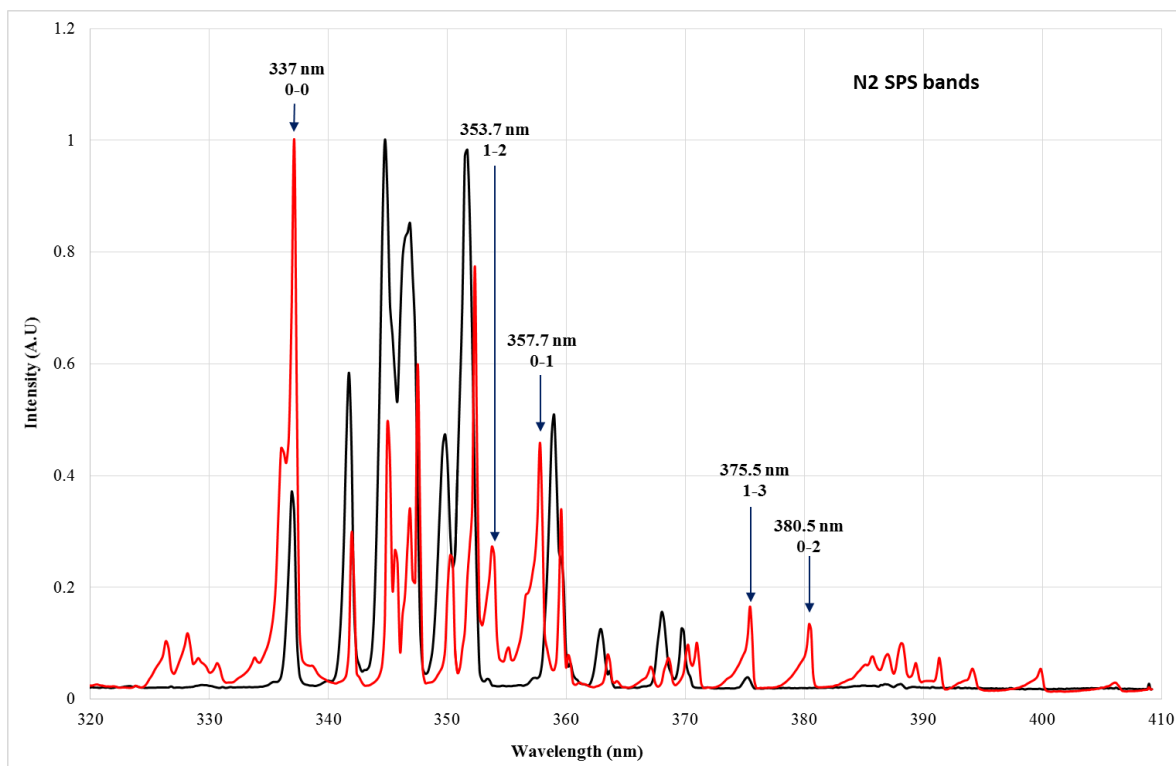

**Figure 3.** Presents a typical UV spectrum of the neon discharge (black curve). Only Ne I lines listed in the NIST data base are observed. The red curve shows the same UV spectrum when plasma reaches the entrance of the reactor and the N$_2$ SPS bands are visible (28).

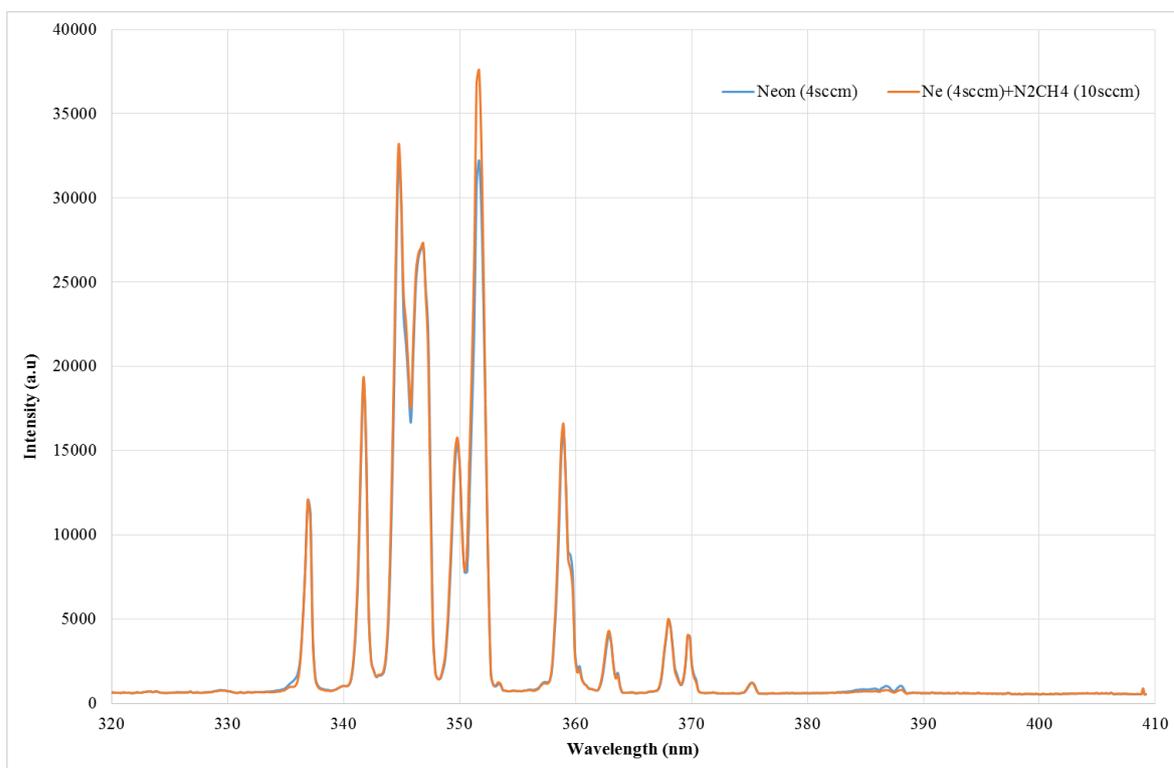

**Figure 4.** UV spectrum with neon in the discharge tube (blue) and neon facing APSIS filled with the N$_2$-CH$_4$ mixture (orange). The fact that the two spectra are exactly the same shows that N$_2$ is not excited by the discharge.

A microwave plasma source for VUV atmospheric photochemistry

This is why the surfatron is pushed back from the entrance of the reactor until the end of the discharge is 5cm away from it. Figure 4 compares UV spectra of the only neon source and the APSIS reactor filled with $N_2$-$CH_4$ and irradiated by the surfatron source in its new position. $N_2$ emission bands are no more observed, even for the maximum (10 sccm) gas flow of $N_2$-$CH_4$. From that we can conclude that $N_2$ is not excited by electrons, and that there is no electron-driven chemistry in the reactor with this position of the surfatron.

*1.1.2 Preventing neon-metastable-atom-driven chemistry.* Another possible source of energy could also be the neon metastable states $^3P^0_{2-0}$ at 16.61 and 16.71 eV. Those metastable atoms are a potential source of energy for ionizing molecular nitrogen as its ionization threshold is at 15.58 eV; this is why it is crucial to determine if they can reach the reactor. As metastable atoms are embattled in the neon flow, their diffusion characteristic time $\tau_D$ has to be compared with their travel time $\tau_t$ between the end of the discharge and the reactor entrance.

In our experimental conditions, the depopulation is mainly done by radial diffusion to the discharge which means that the metastable atom density depends on the Bessel function of the first kind $J_0$ (19):

$$n(r) = n_0 \times J_0(\frac{2.405 \times r}{R}) \qquad (3)$$

The inverse of the diffusion characteristic time is then $\frac{1}{\tau_D} = \frac{D}{\Lambda^2}$ where $\Lambda$, the diffusion length, equals $\frac{R}{2.405}$ where R is the radius of the tube. The neon metastable diffusion coefficient is D=170 cm².s⁻¹ at 1 Torr and 300K (20). In our working conditions, (P=1 mbar, T=300K, R=4mm), the diffusion time of the neon metastable atoms is $\tau_D$=1.6×10⁻⁴ s.

For a neon gas flow Q=4sccm, the velocity is V=1.3×10² cm.s⁻¹ and the travel time on this 5cm-length is $\tau_t$ =4×10⁻² s. As $\tau_t \gg \tau_D$, the neon metastable atoms are destroyed before arriving to the reactive zone.

We are now sure that the only energy input from our lamp are the VUV photons, making it fully compatible to the kind of experiments we intend to perform.

*1.1.3 Discharge length.* In order to increase the VUV photon flux, one can increase the micro-wave power deliver to the discharge, but this will modify the discharge length, which could then enter the reactor. This is why we measured the length of the discharge downstream the surfatron-device as a function of pressure in different conditions of power (Figure 5). The length increases with the pressure. Above 1 mbar, it depends only on the micro-wave power with a maximum of 18 cm at a pressure of 1.7 mbar and a power of 80W.

Then, for further use for photochemistry, we know where the surfatron must be positioned in order to have the maximum of VUV flux without any induced plasma reactions whatever the pressure and/or the power are.

A microwave plasma source for VUV atmospheric photochemistry

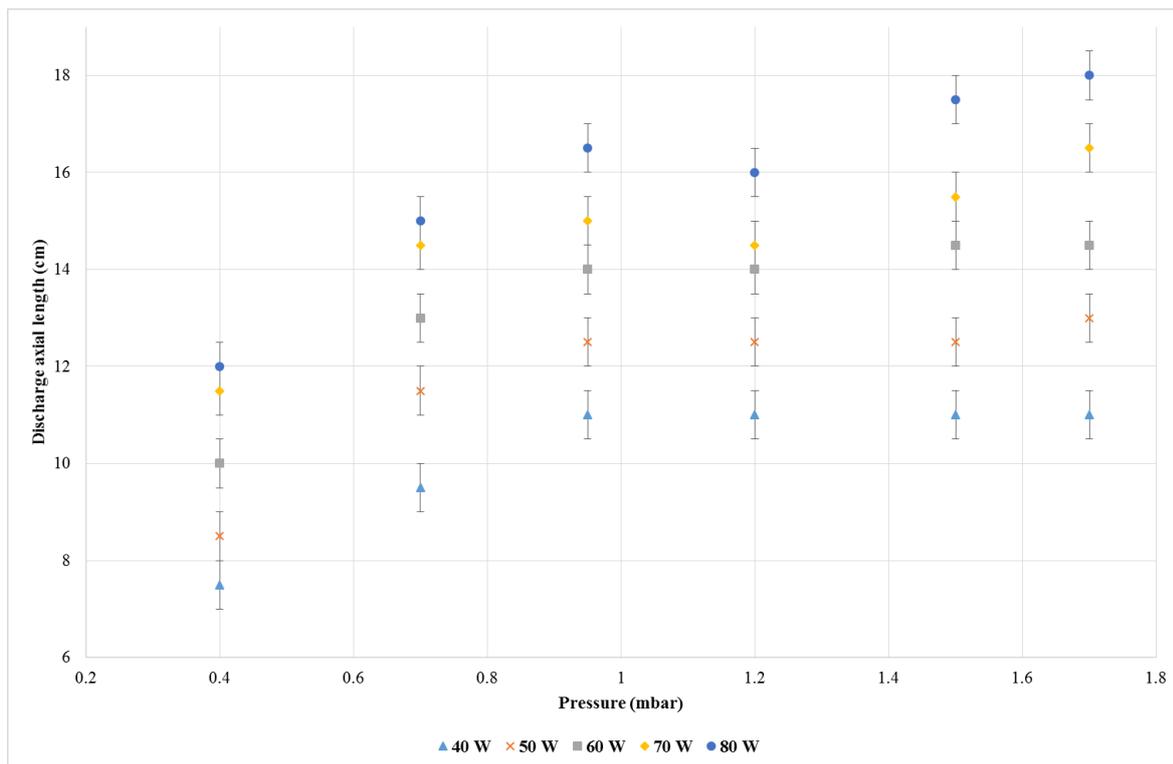

**Figure 5:** Length of the neon microwave discharge (upstream the surfatron device) for different conditions of pressure and power.

*2.2 Characterization of the VUV source*

In order to fully characterize the source in the VUV range, the intensities of the two neon resonant lines (Table 1) are recorded in different conditions of pressure and power for the discharge.

Typical spectra recorded for the two resonance lines of neon are presented on Figure 6. We notice that the two lines present different behaviours regarding the pressure, this is why we will study them separately.

In order to calculate the VUV flux emitted by the surfatron source, we need the spectral responses of both the grating and the photodetector, without forgetting the absorption of the neon present inside the monochromator as described in section 1.2.

A microwave plasma source for VUV atmospheric photochemistry

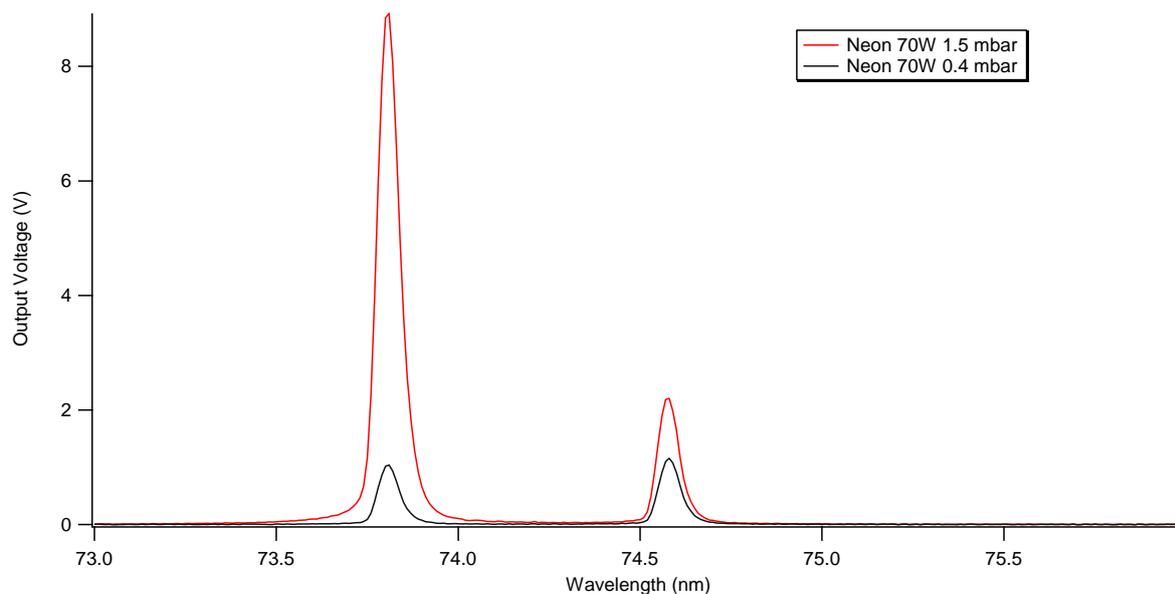

**Figure 6.** The two resonant emission lines of Ne I at 73.6 and 74.3 nm at 70W for two different pressures showing different intensity ratio between the two lines.

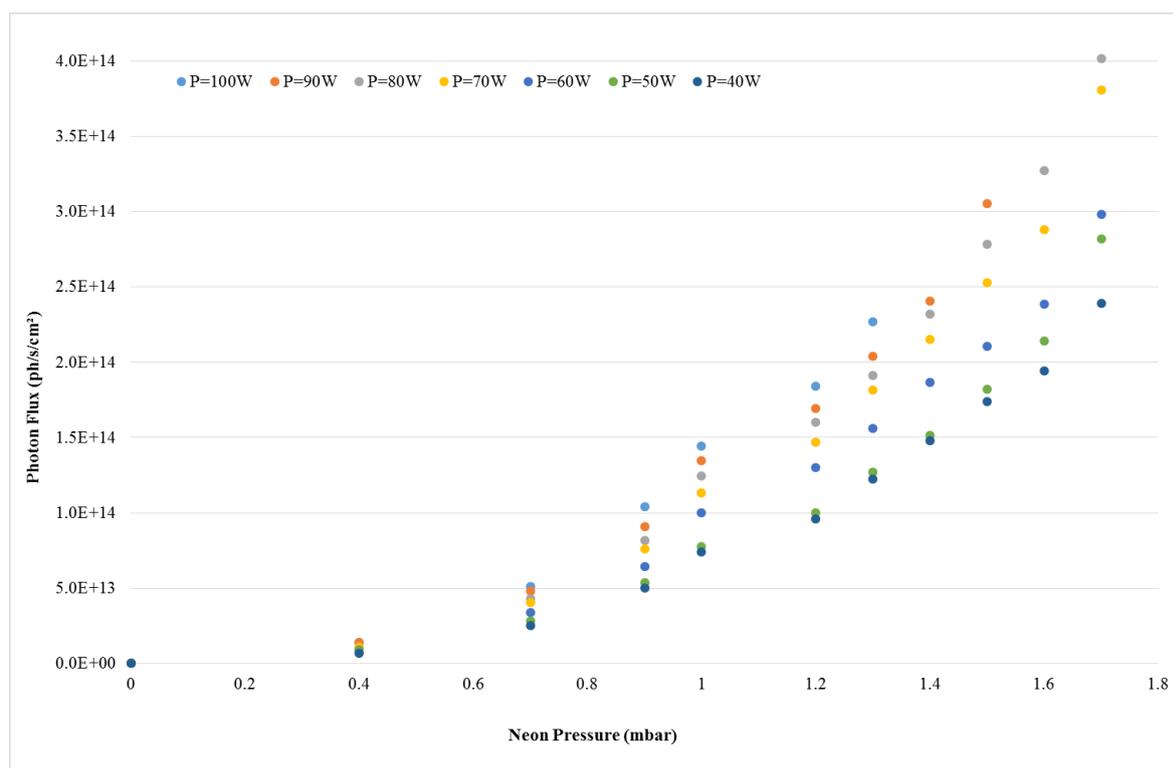

**Figure 7.** Photon flux emitted by the neon microwave discharge in different conditions of power versus the pressure. At high powers ($P_w$>70W) and high pressures (P >1.5 mbar), the amplifier saturates this is why the data are not shown on this graph.

The calculations have been performed first on the 73.6 nm line for different gas pressures and microwave powers (Figure 7; Figure 8). The order of magnitude of the source photon flux is $10^{14}$ ph.s$^{-1}$.cm$^{-2}$, with a maximum of $4\times10^{14}$ ph.s$^{-1}$.cm$^{-2}$ (1.7 mbar and 100W) and a minimum of $2\times10^{13}$ ph.s$^{-1}$.cm$^{-2}$ (0.4 mbar and 40W).

A microwave plasma source for VUV atmospheric photochemistry

Figure 7 shows how the photon flux of the lamp varies versus the pressure for different values of power. It reveals a general profile of quadratic growth regarding the neon pressure as fitted on the orange curve on Figure 9.

On Figure 8, the variations of the photon flux versus the power delivered into the discharge are presented, for different values of pressures. Here, the growth looks linear.

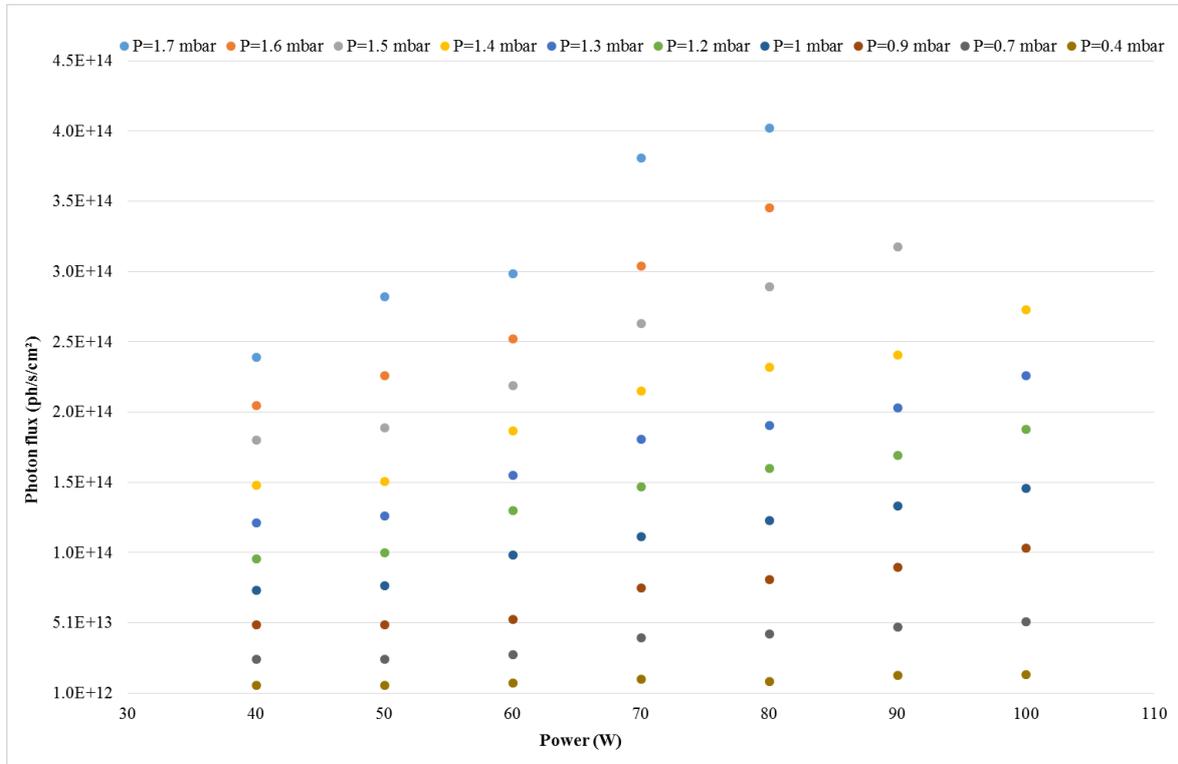

**Figure 8.** Photon flux emitted by the neon microwave discharge in different conditions of pressure versus the power.

These two trends versus the pressure and the power can be explained by a collisional radiative model. The intensity of the discharge is directly proportional to the electron density $n_e$, the neon concentration in its fundamental state [Ne], $\nu_L$ that is the inverse of the lifetime of the radiative level and an excitation coefficient k function of the electronic temperature:

$$I(73.6nm) \propto \frac{n_e[Ne]k(T_e)}{\nu_L} \quad (4)$$

In our pressure range, we assume that the electron energy remains practically constant which means that the ionization degree is supposed to stay constant as well. For a given power, arising the gas pressure would then directly increase the neon density [Ne] but also the electron density $n_e$ which explains the quadratic law of the 73.6 nm-line intensity. Moreover, at a fixed pressure, increasing the delivered power would here have a linear impact on the electron density $n_e$, while the electronic temperature $T_e$ and the neon concentration remain constant.

We can now focus on the evolution of the 74.3nm-line.

When the 74.3nm-photon flux is plotted versus the pressure for an arbitrary power of 70W, it looks like it does not follow the same quadratic trend as the 73.6 nm one (Figure 9). In fact, its intensity varies very linearly with the pressure which means that another phenomenon counterbalances the expected quadratic rise of the photon flux.

A microwave plasma source for VUV atmospheric photochemistry

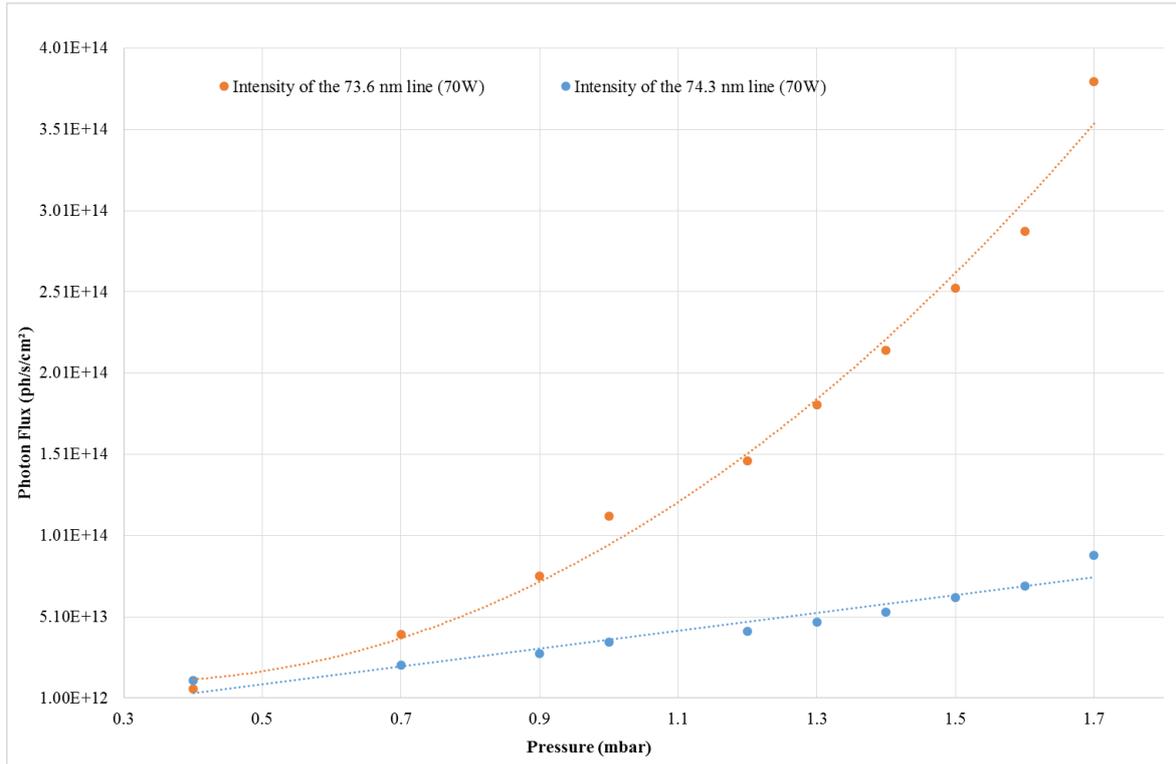

**Figure 9.** Compared photon flux evolution, in function of the pressure, of the two neon emission lines at 73.6 and 74.3 nm for an arbitrary given power of 70W and their quadratic and linear fits (respectively).

In addition to the depopulation of this energy level by radiative photon emission, one lead is to involve a quenching effect. The lifetime of the radiative level $^3P_1$, that emits the 74.3 nm line, was measured by (21) in parallel to the two metastable states $^3P_2$ and $^3P_0$ (Figure 10). These lifetimes are quite similar (few ms). As the 74.3nm-line is resonant, its level $^3P_1$ is re-populated and its density is on the same order of magnitude than the metastable states as (22) has measured in a neon microwave discharge.

The quenching coefficient of the $^3P_1$ radiative level (at 16.67eV) due to the transfer towards the $^3P_2$ metastable state, is $k_u = 4.2 \times 10^{-14}$ cm$^3$s$^{-1}$ (21). The efficiency of this quenching is related to the small energy gap of $5.17 \times 10^{-2}$ eV between those two states. After collision, this energy is shared between the two atoms and corresponds to a temperature of approximately 300K, which is more or less the gas temperature in the discharge.

Equation (5) gives the intensity of the 74.3nm-line taking into account this potential quenching effect. In this equation, $\nu_L$ corresponds to the inverse of the $^3P_1$ level lifetime. Then, the quenching seems to be dominant here, this is why the 74.3nm-line intensity increases linearly with the electron density and thus the pressure.

$$I(74.3\ nm) \propto \frac{n_e[Ne]k(T_e)}{\nu_L + k_Q[Ne]} \tag{5}$$

A microwave plasma source for VUV atmospheric photochemistry

For the 73.6 nm line, which corresponds to the $^1P_1$ level, the quenching is less efficient, because the transfer to the $^3P_0$ metastable state presents a higher energy gap of 0.133 eV i.e a temperature of 770 K for each atom, too hot for our experimental conditions. This level is then mainly depopulated by radiative transfer.

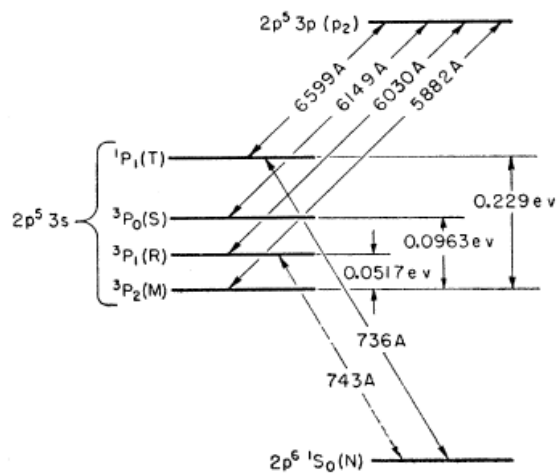

**Figure 10**. Gotrian diagram of the neon energy levels taken from (21)

To conclude, the pressure at which we are working will determine which line(s) will contribute to our photochemistry experiment. For high pressures (P>0.7 mbar), the radiative emission from the 73.6-nm line is clearly dominant compared to the steady contribution of the 74.33-nm one. However, when going low in pressure the contribution of both lines have to be taken into account.

*2.3  First photochemistry experiment*

When the source is coupled to the APSIS reactor, an $N_2$-$CH_4$ mixture (95% $N_2$, 5% $CH_4$) is injected. First, 4 sccm of neon are injected inside the discharge tube, which corresponds to a discharge pressure P=1.3 mbar. In a second time, in order not to contaminate the discharge, we inject 2 sccm of $N_2$-$CH_4$ in APSIS for a pressure of 0.9 mbar. This neon flow permits a relevant photon flux for the VUV source but also prevent the reactive $N_2$-$CH_4$ mixture from entering the discharge tube, this is why we chose it for our first photochemistry experiment. The power delivered to the discharge is $P_w$=80W which corresponds to a photon flux of $1.9 \times 10^{14}$ ph.s$^{-1}$.cm² or $0.95 \times 10^{14}$ ph.s$^{-1}$ for the tube section of 0.5cm².

Two effects have to be monitored by mass spectrometry: first, if the reactive signals are getting lower through the experiment, as it would mean that there is an effective consumption due to the photon interaction; and, in a second time, the signals of some hydrocarbon and/or N-bearing species (Titan-like molecules) are followed in the expectation of seeing a production. Here, we focus on the mass m/z=15 in order to represent the $CH_4$ reactive, through its $CH_3^+$ fragment, because the real $CH_4$ mass m/z=16 can overlap multiple species fragments. In addition, we look at some Titan-like molecules: m/z=27 for

A microwave plasma source for VUV atmospheric photochemistry

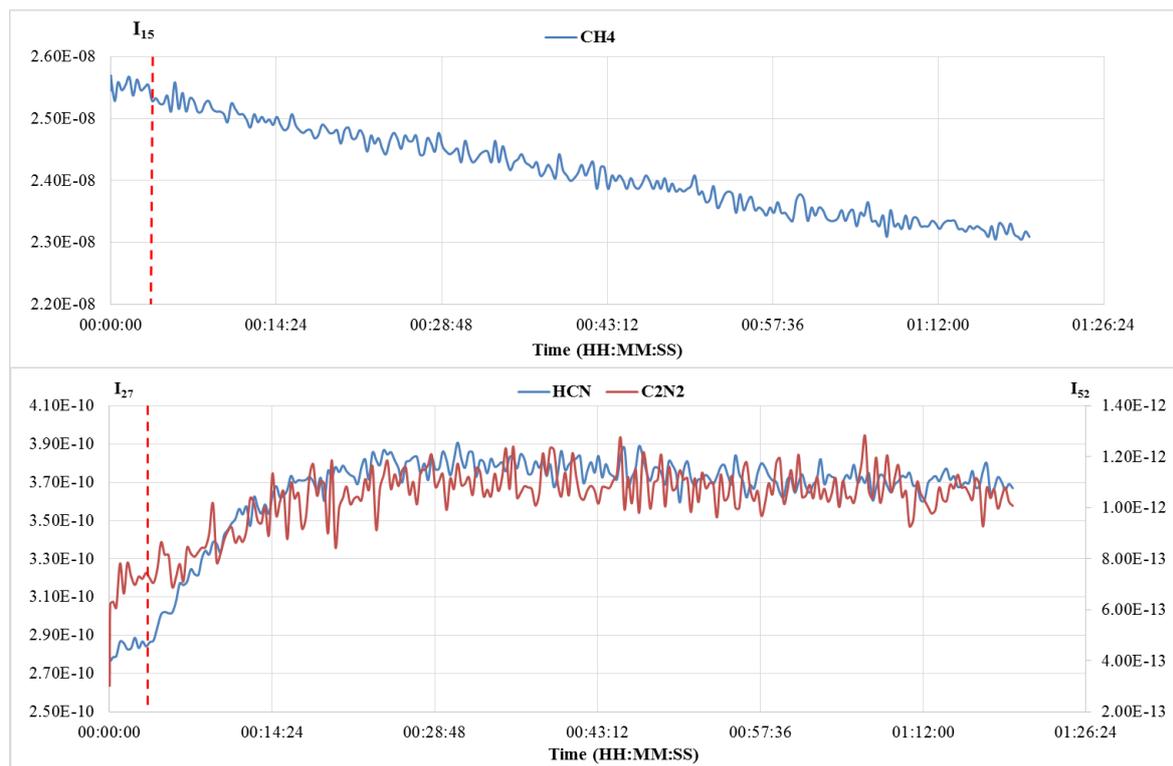

**Figure 11.** Time-monitoring mass spectrum in MID mode following the consumption of the reactive $CH_4$ (top) and the production of the HCN and $C_2N_2$ molecules (bottom). The red dotted lines mark the beginning of the irradiation of the APSIS reactor with the VUV source.

HCN and m/z=52 for $C_2N_2$. On Figure 11, the red dotted line marks the moment when the VUV source is turned ON. The signals of the selected products immediately start to increase, while the m/z=15 signal decreases. After approximately 20 minutes, a quasi-stationary is reached for the products.

The presence of those molecules highlights the incorporation of both C and N atoms coming from the expected dissociation of $N_2$ and $CH_4$. Methane has been identified as a key reactant to initiate efficient organic growth, in contrast to the other major form of carbon commonly found in planetary atmospheres, carbon dioxide $CO_2$. However, methane chemistry leads to the only production of complex hydrocarbon molecules, with no heavy heteroatoms of interest for prebiotic chemistry such as nitrogen. (23). $N_2$ photolysis leads to reactive forms of nitrogen, atomic or ionized, which react with hydrocarbons to produce nitrogen containing organic compounds. These nitrogen containing species have a strong interest for prebiotic chemistry (24). The most abundant gas-phase nitrogen-bearing products in Titan atmosphere are nitriles molecules (R-CN) as the HCN and $C_2N_2$ molecules that we found in our experiment. The goal of the photochemistry experiment is then achieved, which ends to validate the suitability of the VUV source.

## 3. DISCUSSION

For our atmospheric applications, we compare those data to the solar flux. As a matter of fact, at Titan the solar flux is of $10^7$ ph.s$^{-1}$.cm$^{-2}$ at 75 nm. This means that our $N_2$-$CH_4$ reactive medium receives approximately $10^7$ more 75-nm-photons making the chemical reactions more efficient in terms of kinetics.

It is interesting to compare the photon flux numbers to the ones of the VUV DESIRS beamline at the synchrotron SOLEIL (France) which offers a wide energy range between 5 and 40 eV (or 31 to 248 nm). At approximately 17 eV (~75 nm) and for a resolving power of 1000, between $7\times10^{12}$ ph.s$^{-1}$.cm$^{-2}$ for a 4mm×8mm spot and $10^{16}$ ph.s$^{-1}$.cm$^{-2}$ for a 200 µm×100µm spot are delivered (15).

Despite providing only one specific wavelength at a time, our source is then competitive to a VUV

A microwave plasma source for VUV atmospheric photochemistry synchrotron beamline regarding the photon flux as it goes from $2\times10^{13}$ ph.s$^{-1}$.cm$^{-2}$ up to $4\times10^{14}$ ph.s$^{-1}$.cm$^{-2}$.

Regarding the pressures, the one in the photochemical reactor is higher than in Titan's upper atmosphere but this prevents any undesired wall effect (18). Moreover, our setup does not requires a differential pumping system between the source and the 1 mbar-reactor; which is unavoidable at a synchrotron beamline where the source is under ultra-high vacuum at $10^{-8}$ mbar.

*4.* **Conclusion**

So, in conclusion, VUV plasma source based on surfatron-driven microwave discharges have proven to be an efficient and reliable tool for VUV atmospheric photochemistry experiments working at low pressure. For the ones using neon, it has been demonstrated here that the only energy source injected inside the reactive medium is the VUV photons from its two resonant lines (73.6 and 74.3 nm). Other sources are indeed possible (neon metastable atoms and electrons from the discharge) but the system has been adapted so that they will not disturb the desired pure photochemistry. In this case, the key parameter appears to be the distance between the end of the discharge and the entrance of the photochemical reactor. Also, always for neon, the behaviours of the two resonant wavelengths have been investigated in order to characterize their contribution in several conditions of pressure and power. The 73.6-nm line is largely dominant in our working conditions, but its intensity decreases with the pressure (quadratic law) and the power (linear law). This behaviour is different from the 74.3-nm line which remains quite stable whatever the pressure or the power are, possibly because of a quenching effect. It results that at low pressure (<0.7 mbar) the 74.3-nm one becomes dominant. To further investigate these relative intensities, a complete modelling of the discharge in our conditions taking into account the electron energy distribution function would be required. In fact, according to (25) and their model with a 10Torr- DC discharge, the 74.3-nm line has a higher intensity than the 73.6-nm one, which is not in agreement with our experimental results.

Moreover, at 73.6nm, the order of magnitude for the photon flux is $10^{14}$ ph.s$^{-1}$.cm$^{-2}$, which seems largely comparable to a VUV synchrotron beamline at specific wavelength. The working conditions of the surfatron-based VUV source are quite flexible (regarding the pressures and power), allowing the photon flux to vary from $2\times10^{13}$ ph.s$^{-1}$.cm$^{-2}$ up to $4\times10^{14}$ ph.s$^{-1}$.cm$^{-2}$ and offer measurable photochemistry results. Our new setup, without any window separating the source and the photoreactor, can change the way atmospheric photochemistry experiments have been performed so far, especially for the ones focusing of nitrogen-dominated atmospheres such as the Earth, the Early Earth, Titan or Pluto.

More specifically, our setup showed its ability to simulate the formation of nitriles in Titan's atmosphere. Those are not correctly predicted by the current photochemical models (26). Even the simplest and most abundant nitrile, HCN, is found to be predicted by neutral photochemistry with concentrations three to ten times larger than measured by the Cassini space mission (27). Such disagreements between model predictions and observations reveal a poor knowledge of the chemistry involving nitrile in general and HCN in particular. Without the constraint of adapting the system to high pressure gradients between the source and the reactor or the low availability of synchrotron beamlines, our experiment will provide the opportunity to explore the chemistry of nitriles and improve our knowledge in this area.

*5.* **Acknowledgments.**

S. Tigrine would like to thank the IDEX Paris Saclay for her Phd thesis grant and N. Carrasco thanks the ERC Starting Grant PRIMCHEM, grant agreement n°636829. The Centre National d'Etudes Spatiales (CNES) supported the design of the APSIS reactor. All the authors also wish to thank Dr L. Nahon for his kind support and advices.

*6.* **References**

1. Moisan M, Beaudry C, Lepprince P. A new HF device for the production of long plasma columns at a high electron density. Phys Lett A [Internet]. 1974;50(2):125–6. Available from:

A microwave plasma source for VUV atmospheric photochemistry

http://www.sciencedirect.com/science/article/pii/0375960174909037

2. Moisan M, Beaudry C, Leprince P. A Small Microwave Plasma Source for Long Column Production without Magnetic Field. Vol. 3, IEEE Transactions on Plasma Science. 1975. p. 55–9.

3. Moisan M, Zakrzewski Z. Plasma sources based on the propagation of electromagnetic surface waves. J Phys D Appl Phys [Internet]. 1991;24(7):1025. Available from: http://stacks.iop.org/0022-3727/24/i=7/a=001

4. Espinho S, Felizardo E, Tatarova E, Dias FM, Ferreira CM. Vacuum ultraviolet emission from microwave Ar-H2 plasmas. Appl Phys Lett [Internet]. 2013;102(11):1–5. Available from: http://scitation.aip.org/content/aip/journal/apl/102/11/10.1063/1.4796134

5. Northway MJ, Jayne JT, Toohey DW, Canagaratna MR, Trimborn A, Akiyama KI, et al. Demonstration of a VUV Lamp Photoionization Source for Improved Organic Speciation in an Aerosol Mass Spectrometer. Aerosol Sci Technol [Internet]. 2007;41(9):828–39. Available from: http://www.tandfonline.com/doi/abs/10.1080/02786820701496587

6. Hnilica J, Kudrle V, Potocnakova L. Surface treatment by atmospheric-pressure surfatron jet. IEEE Trans Plasma Sci. 2012;40(11 PART1):2925–30.

7. Waite JH, Young DT, Cravens TE, Coates a J, Crary FJ, Magee B, et al. The process of tholin formation in Titan's upper atmosphere. Science [Internet]. 2007;316(5826):870–5. Available from: http://www.ncbi.nlm.nih.gov/pubmed/17495166\nhttp://www.sciencemag.org/cgi/doi/10.1126/science.1139727

8. Clarke DW, Joseph JC, Ferris JP. The Design and Use of a Photochemical Flow Reactor: A LaboratoryStudy of the Atmospheric Chemistry of Cyanoacetylene on Titan. Icarus. 2000;147:282–91.

9. Thuillier G, Floyd L, Woods TN, Cebula R, Hilsenrath E, Hersé M, et al. Solar irradiance reference spectra for two solar active levels. Adv Sp Res. 2004;34(2):256–61.

10. Ádámkovics M. Photochemical formation rates of organic aerosols through time-resolved in situ laboratory measurements. J Geophys Res [Internet]. 2003;108(E8):5092. Available from: http://doi.wiley.com/10.1029/2002JE002028

11. Yoon YH, Hörst SM, Hicks RK, Li R, de Gouw JA, Tolbert MA. The role of benzene photolysis in Titan haze formation. Icarus [Internet]. Elsevier Inc.; 2014;233:233–41. Available from: http://dx.doi.org/10.1016/j.icarus.2014.02.006

12. Es-sebbar E, Bénilan Y, Fray N, Cottin H, Jolly A, Gazeau M-C. Optimization of a Solar Simulator for Planetary-Photochemical Studies. Astrophys J Suppl Ser [Internet]. 2015;218(2):19. Available from: http://stacks.iop.org/0067-0049/218/i=2/a=19?key=crossref.2aa362460a105932a493d88f955dbbb6

13. Imanaka H, Smith MA. Formation of nitrogenated organic aerosols in the Titan upper atmosphere: SI. Proc Natl Acad Sci U S A [Internet]. 2010;107(28):12423–8. Available from: http://www.pnas.org/content/107/28/12423.short

14. Peng Z, Gautier T, Carrasco N, Pernot P, Giuliani A, Mahjoub A, et al. Titan's atmosphere simulation experiment using continuum UV-VUV synchrotron radiation. J Geophys Res Planets [Internet]. 2013;118(4):778–88. Available from: http://dx.doi.org/10.1002/jgre.20064

15. Nahon L, De Oliveira N, Garcia GA, Gil JF, Pilette B, Marcouillé O, et al. DESIRS: A state-of-the-art VUV beamline featuring high resolution and variable polarization for spectroscopy and dichroism at SOLEIL. J Synchrotron Radiat. 2012;19(4):508–20.

16. NIST Database [Internet]. Available from: http://physics.nist.gov/PhysRefData/ASD/lines_form.html

17. Hébrard E, Marty B. Coupled noble gas-hydrocarbon evolution of the early Earth atmosphere upon solar UV irradiation. Earth Planet Sci Lett [Internet]. Elsevier B.V.; 2014;385:40–8. Available from: http://dx.doi.org/10.1016/j.epsl.2013.10.022

18. Carrasco N, Giuliani A, Correia JJ, Cernogora G. VUV photochemistry simulation of planetary upper atmosphere using synchrotron radiation. J Synchrotron Radiat. 2013;20(4):587–9.

19. Moisan M, Pelletier J. Physique des plasmas collisionnels (Application aux dcharges hautes frquences). EDP sciences; 2006.

20. Dixon JR, Grant FA. Decay of the triplet p levels of neon. Phys Rev. APS; 1957;107(1):118.

21. Phelps A V. Diffusion, de-excitation, and three-body collision coefficients for excited neon atoms. Phys Rev. 1959;114(4):1011–25.

22. Ricard A, Barbeau C, Besner A, Hubert J, Margotchaker J, Moisan M, et al. Production of Metastable and Resonant Atoms in Rare-Gas (He, Ne, Ar) Radio-Frequency and Microwave-Sustained Discharges. Can J Phys [Internet]. 1988;66(8):740–8. Available from: <Go to ISI>://A1988R027400015

23. Raulin F, Brasse C, Poch O, Coll P. Prebiotic-like chemistry on Titan. Chem Soc Rev [Internet]. 2012;41(16):5380–93. Available from: http://dx.doi.org/10.1039/C2CS35014A

24. Dutuit O, Carrasco N, Thissen R, Vuitton V, Alcaraz C, Pernot P, et al. CRITICAL REVIEW OF N, N + , N + 2 , N ++ , And N ++ 2 MAIN PRODUCTION PROCESSES AND REACTIONS OF RELEVANCE TO TITAN'S ATMOSPHERE. Astrophys J Suppl Ser [Internet]. 2013;204(2):20. Available from: http://stacks.iop.org/0067-0049/204/i=2/a=20?key=crossref.23de8c3395d21023ed686fde089380fb

25. Zissis G, Damelincourt JJ. Influence du mode d'alimentation sur la production du rayonnement VUV des raies de résonance de Ne dans une décharge électrique à basse pression. Le J Phys IV. EDP sciences; 1999;9(PR5):Pr5–27.

26. Loison JC, Hébrard E, Dobrijevic M, Hickson KM, Caralp F, Hue V, et al. The neutral photochemistry of nitriles, amines and imines in the atmosphere of Titan. Icarus [Internet]. Elsevier Inc.; 2015;247:218–47. Available from: http://dx.doi.org/10.1016/j.icarus.2014.09.039

27. Hébrard E, Dobrijevic M, Loison JC, Bergeat a., Hickson KM. Neutral production of hydrogen isocyanide (HNC) and hydrogen cyanide (HCN) in Titan's upper atmosphere. Astron Astrophys. 2012;541:A21.

28. Lofthus A, Krupenie PH. The spectrum of molecular nitrogen. J Phys Chem Ref Data. AIP Publishing; 1977;6(1):113–307.